\documentclass[notitlepage,nofootinbib,preprintnumbers,amssymb,superscriptaddress,
twocolumn
]{revtex4-2}
\usepackage{amsfonts,amssymb,amsmath,mathtools,graphicx,color,bm,orcidlink}
\definecolor{ultramarine}{rgb}{0.07, 0.04, 0.56}
\definecolor{teal}{rgb}{0.0, 0.5, 0.5}
\definecolor{indigo(dye)}{rgb}{0.0, 0.25, 0.42}
\usepackage{hyperref}
\hypersetup{
colorlinks=true,
citecolor=ultramarine,
linkcolor=teal,
urlcolor=indigo(dye),
}

\let\originalleft\left
\let\originalright\right
\renewcommand{\left}{\mathopen{}\mathclose\bgroup\originalleft}
\renewcommand{\right}{\aftergroup\egroup\originalright}
\usepackage{autobreak}

\newcommand{\na}{\nabla}

\newcommand{\be}{\begin{equation}}  
\newcommand{\ee}{\end{equation}}
\newcommand{\bem}{\begin{bmatrix}}
\newcommand{\eem}{\end{bmatrix}}

\newcommand{\mn}{{\mu \nu}}

\allowdisplaybreaks

\begin{document}

\preprint{RESCEU-25/26}

\title{Avoided crossings and resonances in black hole ringdown with coupled fields:\\
A case study of the Einstein--Maxwell--axion system}

\author{Takuya Takahashi\,\orcidlink{0000-0002-4894-6108}}
\email{takuya.takahashi@resceu.s.u-tokyo.ac.jp}
\affiliation{Research Center for the Early Universe (RESCEU), Graduate School of Science, The University of Tokyo, Tokyo 113-0033, Japan}

\author{Hayato Motohashi\,\orcidlink{0000-0002-4330-7024}}
\affiliation{Department of Physics, Tokyo Metropolitan University, 1-1 Minami-Osawa, Hachioji, Tokyo 192-0397, Japan}

\author{Kazufumi Takahashi\,\orcidlink{0000-0002-4070-1675}}
\affiliation{Research Center for the Early Universe (RESCEU), Graduate School of Science, The University of Tokyo, Tokyo 113-0033, Japan}

\begin{abstract}
Avoided crossings in black hole quasinormal-mode (QNM) spectra exhibit characteristic resonant properties, including enhanced excitation factors. We extend the study of this phenomenon to perturbation systems with multiple coupled fields and investigate its consequences for ringdown waveforms. As a concrete example, we consider a charged black hole in the Einstein--Maxwell--axion system, whose QNM branches can be classified according to the gravitational, electromagnetic, and axion modes to which they continuously connect in the decoupling limit. We focus on an avoided crossing between the gravity- and electromagnetic-led branches thus defined. We verify that the QNM reconstruction accurately reproduces the ringdown portion of the directly evolved time-domain waveform. Near the avoided crossing, an effectively slower decay emerges for particular initial perturbations, while the resonantly enhanced contributions from the two QNMs largely cancel in the physical waveform. The QNM decomposition further reveals an exchange of the physical characters of the two branches across the resonance. These results characterize the imprint of resonances on black hole ringdown in systems with multiple coupled fields.
\end{abstract}

\maketitle

\section{Introduction}\label{sec:intro}
Black hole (BH) ringdown provides a particularly clean setting for studying the dynamics of perturbed BHs.
After a compact binary merger, the remnant BH relaxes toward a stationary configuration through damped oscillations described by its quasinormal modes (QNMs)~\cite{Vishveshwara:1970zz,Press:1971wr,Teukolsky:1973ha}.
These modes can carry information not only about the properties of the remnant spacetime but also about the gravitational theory and possible additional degrees of freedom~\cite{Kokkotas:1999bd,Dreyer:2003bv,Berti:2005ys,Berti:2009kk,Konoplya:2011qq,Hatsuda:2021gtn}.
With the rapidly growing number of gravitational wave detections from compact binary mergers~\cite{LIGOScientific:2026sit,LIGOScientific:2026wfs}, ringdown has therefore become an increasingly important probe of fundamental physics, offering opportunities to test gravity and search for new particles.
Alongside these observational developments, the theoretical understanding of BH perturbations and ringdown has also continued to advance.
Despite its long history, the field has seen substantial theoretical progress in recent years~\cite{Berti:2025hly}.

The QNM spectrum of a BH consists of a discrete set of complex frequencies, including the fundamental mode and a sequence of overtones.
An intriguing feature of QNM spectra is the occurrence of avoided crossings, where two QNM frequencies come close to each other without crossing, and associated resonant properties~\cite{Motohashi:2024fwt}. Such avoided crossings are closely associated with exceptional points, at which both the eigenfrequencies and the corresponding eigenmodes coalesce. 
Near an exceptional point, the excitation factors~\cite{Leaver:1986gd,Berti:2006wq}, which characterize the excitation amplitudes of the corresponding QNMs, can exhibit resonant enhancement. 
These spectral and excitation properties are characteristic of resonant phenomena in non-Hermitian physics, providing an interesting connection between BH QNMs and non-Hermitian systems.

An important question is how such resonant behavior manifests itself in the physical ringdown waveform. 
Although the excitation factors are strongly enhanced near the resonance, their enhancements 
occur in a nearly point-symmetric manner in the complex plane, so that they typically largely cancel in the physical waveform, preventing a corresponding resonant enhancement of the waveform amplitude~\cite{Oshita:2025ibu,Kubota:2025hjk}. 
Nevertheless, as the two QNM frequencies become nearly degenerate, their superposition effectively generates a contribution proportional to time, resulting in a characteristic slower decay~\cite{Yang:2025dbn,PanossoMacedo:2025xnf}.
This behavior requires particular care when modeling the ringdown in terms of a conventional superposition of two QNMs. 
Although distinguishing this behavior from an ordinary two-QNM superposition in data analysis can be challenging~\cite{Imafuku:2026rpn}, there are several efficient approaches for extracting nearly degenerate QNMs~\cite{Morisaki:2025gyu,Iida:2026jox,Iida:2026ylf}, and the characteristic time dependence may provide a useful signature of an avoided crossing and resonance in BH ringdown.

Previous studies of avoided crossings and resonances in BH QNMs have mostly focused on systems described by a single perturbation variable, including various BH spacetimes within General Relativity (GR)~\cite{Motohashi:2024fwt,Cavalcante:2024swt,Nakamoto:2026lyo,Cavalcante:2026vgr} and phenomenological models with deformed effective potentials~\cite{Yang:2025dbn,PanossoMacedo:2025xnf,Cao:2025afs,OuldElHadj:2026vym,Wu:2026tfx} (but see also Refs.~\cite{Tahara:2026chs,Cheng:2026gxu,Hu:2025efp}).
In theories beyond GR relevant to BH spectroscopy, however, additional dynamical degrees of freedom are typically present.
Motivated by possible interactions between gravitational perturbations and such additional fields, it is therefore important to understand how avoided crossings manifest themselves in BH perturbation theory with multiple coupled fields.

In our previous work~\cite{Takahashi:2025uwo}, we developed a general framework for describing QNM contributions to ringdown waveforms in BH perturbation theory with multiple coupled fields. 
As a concrete example, we applied the formalism to a charged BH in Einstein--Maxwell--axion (EMA) theory.
In the parity-odd sector of this system, the gravitational, electromagnetic, and axion perturbations are coupled in general~\cite{Boskovic:2018lkj}. 
We found an avoided crossing between the QNMs associated with the gravitational and electromagnetic degrees of freedom, accompanied by a resonant enhancement of their excitation factors. 
Interestingly, this resonance between different degrees of freedom appeared across multiple overtones.

In this work, we extend the previous analysis to the time domain and investigate how avoided crossings in a multi-field system affect the physical ringdown waveform.
We first compare the waveform obtained by direct integration of the time-domain perturbation equations with that reconstructed from a superposition of QNMs. 
We then examine how the waveform depends on the system parameters and the choice of initial perturbations.

The rest of this paper is organized as follows.
In Sec.~\ref{sec:general}, we develop a general framework for describing BH ringdown in systems with multiple coupled fields.
In Sec.~\ref{sec:EMA}, we describe BH perturbations in the EMA system, which serves as an illustrative example to which we apply the general formalism.
In Sec.~\ref{sec:timefreq}, we compare the waveforms obtained from the time- and frequency-domain calculations.
In Sec.~\ref{sec:phenomenology}, we examine the impact of avoided crossings in the QNM spectrum on the ringdown waveform in systems with multiple coupled degrees of freedom.
Finally, we draw our conclusions in Sec.~\ref{sec:conc}.
Throughout this paper, we use the units with $c=G=4\pi\epsilon_0=1$, with $c$ the speed of light in a vacuum, $G$ the gravitational constant, and $\epsilon_0$ the permittivity in a vacuum.

\section{General formalism}\label{sec:general}
In this section, we present the general framework of BH perturbation theory for ringdown waveforms in systems with multiple coupled fields.
We consider an $N$-component vector~$\vec{\Psi}(t,r)=(\Psi_1,\cdots,\Psi_N)^{\rm t}$ satisfying the following wave equation:
\begin{align}\label{eq:Peq_t}
    \left(-\frac{\partial^2}{\partial t^2}+\frac{\partial^2}{\partial r_\ast^2}\right)\vec{\Psi}(t,r)-\bm{V}(r)\vec{\Psi}(t,r)=0~,
\end{align}
where $\bm{V}(r)$ is an $N\times N$ potential matrix (not necessarily symmetric), and $r_\ast=r_\ast(r)$ is the tortoise coordinate.
For notational simplicity, we use the same symbol for a function irrespective of whether its radial argument is expressed in terms of $r$ or $r_\ast$.
Here, we assume $r_\ast\to-\infty$ as $r\to r_{\rm h}$ and $r_\ast\to\infty$ as $r\to\infty$, where $r_{\rm h}$ denotes the location of the BH event horizon.\footnote{For simplicity, we assume that all the degrees of freedom share the same horizon radius at $r=r_{\rm h}$, as is the case for the EMA system studied in Sec.~\ref{sec:EMA}. Note, however, that different degrees of freedom can propagate at different speeds and hence have different horizon radii in general (see, e.g., Ref.~\cite{Cardoso:2024qie}).}
Defining the Laplace transform of $\vec{\Psi}(t,r)$ as
\begin{align}
    {\cal L}\vec{\Psi}(t,r)\coloneq\vec{\hat{\Psi}}(\omega,r)=\int_{t_0}^\infty dt\ \vec{\Psi}(t,r)e^{i\omega t}~,
\end{align}
the original fields can be written as
\begin{align}\label{eq:Laplace}
    \vec{\Psi}(t,r)=\frac{1}{2\pi}\int_{-\infty+ic}^{\infty+ic}d\omega\ \vec{\hat{\Psi}}(\omega,r)e^{-i\omega t}~,
\end{align}
with $c>0$.
Thus, in the frequency domain, Eq.~\eqref{eq:Peq_t} can be written as
\begin{align}\label{eq:Peq_f}
    \left(\frac{d^2}{dr_\ast^2}+\omega^2\right)\vec{\hat{\Psi}}(\omega,r)-\bm{V}(r)\vec{\hat{\Psi}}(\omega,r)=\vec{S}(\omega,r)~.
\end{align}
Here, the source term~$\vec{S}$ arises from the initial conditions:
\begin{align}\label{eq:source}
    \vec{S}(\omega,r)=e^{i\omega t_0}\left[i\omega\vec{\Psi}(t,r)-\frac{\partial \vec{\Psi}(t,r)}{\partial t}\right]_{t=t_0}~.
\end{align}

QNMs are defined as homogeneous solutions of Eq.~\eqref{eq:Peq_f} that satisfy the physically appropriate boundary conditions, namely, purely ingoing waves at the horizon and purely outgoing waves at infinity~\cite{Pani:2013pma,Chaverra:2016ttw}.
Assuming $\bm{V}(r)\to 0$ for $r\to r_{\rm h}$ and $r\to\infty$ for simplicity, the ``in-mode'' functions, corresponding to purely ingoing waves at the horizon, and the ``up-mode'' functions, corresponding to purely outgoing waves at infinity, are defined by
\begin{align}
    &\vec{\hat{\Psi}}^{({\rm in},\alpha)}\to e^{-i\omega r_\ast}\vec{\delta}_\alpha~, \quad r_\ast\to-\infty~, \label{eq:inmode} \\
    &\vec{\hat{\Psi}}^{({\rm up},\alpha)}\to e^{+i\omega r_\ast}\vec{\delta}_\alpha~, \quad r_\ast\to+\infty~, \label{eq:upmode}
\end{align}
where $\vec{\delta}_\alpha=(\delta_{\alpha1},\delta_{\alpha2},\cdots,\delta_{\alpha N})^{\rm t}$, and $\delta_{\alpha\beta}$ is the Kronecker delta.
For these solutions, the $2N\times 2N$ Wronskian matrix is given by
\begin{align}
    \bm{W}=\begin{pmatrix}\bm{\Psi}^{\rm (in)} & \bm{\Psi}^{\rm (up)} \\
    \partial_{r_\ast}\bm{\Psi}^{\rm (in)} & \partial_{r_\ast}\bm{\Psi}^{\rm (up)}
     \end{pmatrix}~,
\end{align}
where
\begin{align}\label{eq:Psi_matrix}
    \bm{\Psi}^{\rm (in)}=\left(\vec{\hat{\Psi}}^{({\rm in},1)}, \cdots, \vec{\hat{\Psi}}^{({\rm in},N)}\right)~,
\end{align}
while $\bm{\Psi}^{\rm (up)}$ is defined in the same manner.
The determinant of the Wronskian matrix is independent of $r$ and depends only on $\omega$. The QNM frequencies are therefore obtained by solving
\begin{align}
    \det \bm{W}(\omega)=0~.
\end{align}
The QNM frequencies are discrete complex eigenvalues, conventionally ordered by increasing $|{\rm Im}[\omega]|$ and labeled by the overtone number~$n=0,1,2,\cdots$,  where $n=0$ denotes the fundamental mode.
In coupled systems, however, each overtone generally exhibits multiple branches associated with the different degrees of freedom.

Using the Green's matrix method~\cite{Reid:1931,Sisman:2009mk,OuldElHadj:2024psw} (see Appendix~A of Ref.~\cite{Takahashi:2025uwo} for details), the solution of Eq.~\eqref{eq:Peq_f} satisfying the boundary conditions in Eqs.~\eqref{eq:inmode} and \eqref{eq:upmode} is formally given by
\begin{align}\label{eq:sol}
    \vec{\hat{\Psi}}(\omega,r_\ast)=\int dr_\ast'\ \bm{G}(r_\ast,r_\ast')\vec{S}(r_\ast')~.
\end{align}
Assuming that the observer is located sufficiently far from the source, we focus on the region~$r_\ast\geq r_\ast'$, for which the Green's matrix is given by
\begin{align}
    \bm{G}(r_\ast,r_\ast')=\bm{U}^{\rm L}_{N\times 2N}\bm{W}^{({\rm up})}(r_\ast)\bm{W}^{-1}(r_\ast')\bm{L}^{\rm D}_{2N\times N}~,
\end{align}
where
\begin{align}
    \bm{W}^{({\rm up})}=\bm{W}\begin{pmatrix}
    \bm{O} & \bm{O} \\
    \bm{O} & \bm{I}
    \end{pmatrix}~,
\end{align}
and
\begin{align}
\bm{U}^{\rm L}_{N\times 2N}=\begin{pmatrix} \bm{I} & \bm{O} \end{pmatrix}~, \quad \bm{L}^{\rm D}_{2N\times N}=\begin{pmatrix} \bm{O} \\ \bm{I} \end{pmatrix}~.
\end{align}
Here, $\bm{O}$ and $\bm{I}$ denote the $N\times N$ zero matrix and identity matrix, respectively.
Using
\begin{align}
\bm{U}^{\rm L}_{N\times 2N}\bm{W}^{({\rm up})}= \bm{\Psi}^{\rm (up)}\begin{pmatrix} \bm{O} & \bm{I} \end{pmatrix}\equiv \bm{\Psi}^{\rm (up)}\bm{U}^{\rm R}_{N\times 2N}~,
\end{align}
and 
\begin{align}
    \bm{W}^{-1}=(\det \bm{W})^{-1}\tilde{\bm{W}}~,
\end{align}
where $\tilde{\bm{W}}$ is the adjugate matrix of $\bm{W}$, Eq.~\eqref{eq:sol} can be written as
\begin{align}\label{eq:sol2}
\vec{\hat{\Psi}}(\omega,r_\ast)=\frac{1}{\det\bm{W}}\bm{\Psi}^{({\rm up})}\int dr_\ast'\ \bm{T}(r_\ast')\vec{S}(r_\ast')~,
\end{align}
where we have defined
\begin{align}
    \bm{T}(r_\ast')&\coloneq \bm{U}^{\rm R}_{N\times 2N}\tilde{\bm{W}}(r_\ast')\bm{L}^{\rm D}_{2N\times N}~.
\end{align}

To characterize the asymptotic behavior of the mode functions, we introduce the amplitudes of the ingoing and outgoing waves through
\begin{align}
\vec{\Psi}^{({\rm in},\alpha)}\to
\begin{cases}
e^{-i\omega r_\ast}\vec{\delta}_\alpha~, &r_\ast\to-\infty~, \\
e^{-i\omega r_\ast}\vec{A}^{({\rm in},\alpha)}+e^{+i\omega r_\ast}\vec{A}^{({\rm out},\alpha)}~,  &r_\ast\to+\infty~.
\end{cases}
\end{align}
With this definition, the determinant of the Wronskian matrix can be expressed as
\begin{align}
    \det\bm{W}=(2i\omega)^N\det\bm{A}^{(\rm in)}~,
\end{align}
where
\begin{align}
    \bm{A}^{(\rm in)}=\left(\vec{A}^{({\rm in},1)},\vec{A}^{({\rm in},2)},\cdots,\vec{A}^{({\rm in},N)}\right)~.
\end{align}
Applying the inverse Laplace transform~\eqref{eq:Laplace} to Eq.~\eqref{eq:sol2} yields the waveform in the time domain. Retaining only the contributions from the QNM poles, the solution corresponding to the ringdown signal is given by
\begin{align}\label{eq:Psi_t}
\vec{\Psi}(t,r)=&-\sum_{s,n} A_{sn}e^{-i\omega_{s n}t}\bm{\Psi}^{\rm (up)}\int dr_\ast'\ \bm{T}(r_\ast')\vec{S}(r_\ast') \notag \\
&+{\rm (c.c.)}~,
\end{align}
where (c.c.) denotes the complex conjugate, and we have defined
\begin{align}
    A_{s n}\coloneq \frac{i}{(2i\omega_{s n})^N}\left.\left(\frac{d}{d\omega}\det\bm{A}^{({\rm in})}\right)^{-1}\right|_{\omega=\omega_{s n}}~.
\end{align}
Here, we assume that the system involves parameter(s) that control the coupling among different degrees of freedom and that the QNMs can be grouped into branches according to the degrees of freedom to which they are continuously connected in the decoupling limit.
We label these branches by $s$, so that each QNM is specified by $s$ and the overtone number~$n$.
The coefficient~$A_{s n}$ is related to the excitation factor introduced in Ref.~\cite{Takahashi:2025uwo}, with the normalization of the outgoing wave removed, namely, $B_{s n}=A_{s n}\det\bm{A}^{({\rm out})}$.
All frequency-dependent quantities in each term of the sum are evaluated at the corresponding QNM frequency~$\omega=\omega_{sn}$.
In particular, in the asymptotic region, $\bm{\Psi}^{\rm (up)}\sim e^{i\omega_{sn} r_\ast}\bm{I}$ for each QNM contribution in Eq.~\eqref{eq:Psi_t}, which represents the retarded wave.

\section{Einstein--Maxwell--axion system}\label{sec:EMA}
In the remainder of this paper, we focus on the Einstein--Maxwell--axion (EMA) system. This provides an explicit example of the formalism developed in the previous section and illustrates the avoided crossings and resonances between QNM branches originating from different degrees of freedom.
The action of the EMA system is given by
\begin{align}
    S=\frac{1}{4\pi}\int d^4x\sqrt{-g}\biggl[
    \frac{1}{4}R-\frac{1}{4}&F_{\mu\nu} F^{\mu\nu}
    -\frac{1}{2}g^{\mu\nu}\partial_\mu\phi\partial_\nu\phi \notag \\
    &-\frac{1}{4}g_{a\gamma\gamma}\phi F_{\mu\nu}\tilde{F}^{\mu\nu}\biggr]~,
    \label{eq:action_EMA}
\end{align}
where $R$ denotes the Ricci scalar, $F_\mn\coloneq \na_\mu A_\nu-\na_\nu A_\mu$ is the field strength of the electromagnetic field~$A_\mu$, and $\tilde{F}^\mn\coloneq \varepsilon^{\mn\alpha\beta}F_{\alpha\beta}/2$ with the totally antisymmetric tensor~$\varepsilon_{\mn\lambda\sigma}$ defined so that $\varepsilon^{0123}=1/\sqrt{-g}$.
The axion field~$\phi$ is a pseudo-scalar, and $g_{a\gamma\gamma}$ denotes the coupling constant between the electromagnetic and axion fields. Although a potential term can be included in the action, we neglect it throughout this work for simplicity.
We also normalize $\phi$ such that the ratios among the coefficients are rational.

We now consider a static, spherically symmetric, electrically charged BH solution.
When the magnetic charge is absent, regularity at the horizon allows only a constant background axion configuration, which we set to $\phi^{(0)}=0$.
The spacetime is therefore described by the Reissner--Nordstr\"{o}m (RN) metric~\cite{Reissner:1916cle},
\begin{align}
    g^{(0)}_{\mu\nu}dx^\mu dx^\nu=-f(r)dt^2+\frac{1}{f(r)}dr^2+r^2
(d\theta^2+\sin^2\theta\,d\varphi^2)~,
\end{align}
where $f(r)=1-2M/r+Q^2/r^2$ with $M$ and $Q$ the BH mass and charge, respectively, and the EM field is given by $A^{(0)}_\mu dx^\mu=-(Q/r) dt$.
For $0<|Q|<M$, the outer and inner horizons are located at
\begin{align}
r_\pm=M\pm\sqrt{M^2-Q^2}~,
\end{align}
where $r_+$ ($r_-$) denotes the event (Cauchy) horizon.

We linearize the field equations about the above background solution by introducing the perturbations~$g_{\mu\nu}=g^{(0)}_{\mu\nu}+h_{\mu\nu}$, $A_\mu=A^{(0)}_\mu+\delta A_\mu$, and $\phi=\phi^{(0)}+\delta\phi$.
We also expand the perturbations in tensor, vector, and scalar spherical harmonics. In the Regge--Wheeler gauge~\cite{Regge:1957td}, the metric perturbation is decomposed as
\begin{align}
h_{\mu\nu}=\begin{pmatrix}
H_0^{l}Y^{lm} & H_1^{l}Y^{lm} & h_0^{l}S_{\theta}^{lm} & h_0^{l}S_{\varphi}^{lm} \\
\ast & H_2^{l}Y^{lm} & h_1^{l}S_{\theta}^{lm} & h_1^{l}S_{\varphi}^{lm} \\
\ast & \ast & r^2 K^{l}Y^{lm} & 0 \\
\ast & \ast & \ast & r^2K^{l}\sin^2\theta Y^{lm}
\end{pmatrix}~,
\end{align}
where the asterisks denote symmetric components, $Y^{lm}$ are the scalar spherical harmonics, $S_b^{lm}$ ($b\in\{\theta,\varphi\}$) are the axial vector harmonics, and $H_0$, $H_1$, $H_2$, $h_0$, $h_1$, and $K$ are functions of $(t,r)$.
The electromagnetic perturbation is decomposed as
\begin{align}
\delta A_{\mu}=\begin{pmatrix}
u_{1}^l Y^{lm} \\
u_{2}^l Y^{lm} \\
u_{3}^l Y_b^{lm}
\end{pmatrix}
+\begin{pmatrix}
0 \\
0 \\
u_{4}^l S_b^{lm}
\end{pmatrix}~,
\end{align}
where $Y_b^{lm}$ denote the polar vector harmonics, and $u_{1,2,3,4}$ are functions of $(t,r)$. Using the residual $U(1)$ gauge freedom, one can set $u_3=0$.
The axion perturbation is expanded as
\begin{align}
\delta\phi=\frac{\psi^l(t,r)}{r}Y^{lm}~.
\end{align}
The harmonic indices~$l$ and $m$ are implicitly summed over. Since the perturbation equations on a spherically symmetric background are independent of $m$, we suppress the index~$m$. Furthermore, because modes with different $l$ evolve independently, the index~$l$ will also be omitted hereafter.

The perturbation variables separate into polar and axial sectors according to their parity properties. Since the axion 
is a pseudo-scalar, its perturbation couples only to the axial sector, while the polar sector remains unchanged. In the following, we therefore focus on the axial sector.
For the gravitational perturbation, it is convenient to introduce the following master variable:
\begin{align}
h^{\rm odd}=\frac{2}{(l-1)(l+2)}\left[r\left(r^2 \frac{\partial}{\partial r}\left(\frac{h_0}{r^2}\right)-\frac{\partial}{\partial t}h_1\right)+\frac{4Q}{r}u_4\right]~.
\end{align}
In this sector, the only nonvanishing EM perturbation is $u_4$.
The axial perturbations can then be described by a system of three coupled equations for $(h^{\rm odd},u_4,\psi)$.
Hereafter, we omit the superscript~``odd'' and the subscript~``4'' for simplicity.

The linearized field equations yield the following equations:
\begin{align}
&\left(-\frac{\partial^2}{\partial t^2}+\frac{\partial^2}{\partial r_\ast^2}\right)\psi-V_\psi\psi+S_{\psi u} u=0~, \label{eq:psi_pert} \\
&\left(-\frac{\partial^2}{\partial t^2}+\frac{\partial^2}{\partial r_\ast^2}\right)u -V_u u+S_{u\psi}\psi+S_{uh}h=0~, \label{eq:u_pert} \\
&\left(-\frac{\partial^2}{\partial t^2}+\frac{\partial^2}{\partial r_\ast^2}\right)h -V_h h+S_{hu} u=0~, \label{eq:h_pert}
\end{align}
with
\begin{align}
V_\psi&=f(r)\left(\frac{l(l+1)}{r^2}+\frac{2M}{r^3}-\frac{2Q^2}{r^4}\right)~, \\
S_{\psi u}&=f(r)\frac{l(l+1)g_{a\gamma\gamma}Q}{r^3}~, \label{eq:Spsi} \\
V_u&=f(r)\frac{l (l+1)r^2+4 Q^2}{r^4}~, \\
S_{u\psi}&=f(r)\frac{g_{a\gamma\gamma}Q}{r^3}~, \\
S_{uh}&=f(r)\frac{\left(l^2+l-2\right) Q}{2r^3}~, \\
V_h&=f(r)\left(\frac{\left(l^2+l-3\right)r^2+Q^2}{r^4}+\frac{3f(r)}{r^2}\right)~, \\
S_{hu}&=f(r)\frac{8Q}{r^3}~.
\end{align}
Here, the tortoise coordinate is defined by
\begin{align}
    r_\ast\coloneq r+\frac{r_+^2}{r_+-r_-}\log\left(\frac{r-r_+}{M}\right)-\frac{r_-^2}{r_+-r_-}\log\left(\frac{r-r_-}{M}\right)~.
    \label{eq:tortoise}
\end{align}
Although the gravitational and electromagnetic subsectors can be diagonalized as in Refs.~\cite{Boskovic:2018lkj,Takahashi:2025uwo}, we retain the original variables in order to preserve their physical interpretation.
We also note that the normalization of these master variables remains arbitrary up to constant rescalings.

In what follows, we apply the general formalism developed in Sec.~\ref{sec:general} to the system of Eqs.~\eqref{eq:psi_pert}--\eqref{eq:h_pert}. In the present case, we identify $r_{\rm h}=r_+$ and
\begin{align}
    \vec{\Psi}=(\psi,u,h)^{\rm t}~, \quad
    \bm{V}=\begin{pmatrix}
    V_\psi&-S_{\psi u}&0\\
    -S_{u\psi}&V_u&-S_{uh}\\
    0&-S_{hu}&V_h
    \end{pmatrix}~. \nonumber
\end{align}
The axion and electromagnetic perturbations decouple when $g_{a\gamma\gamma}Q=0$, for which $S_{\psi u}=S_{u\psi}=0$, while the gravitational and electromagnetic perturbations decouple for $Q=0$, where $S_{uh}=S_{hu}=0$.

\section{Analysis of time-domain and frequency-domain methods}\label{sec:timefreq}
In this section, we apply the formalism developed in Sec.~\ref{sec:general} to the EMA system described by the perturbation variables~$\vec{\Psi}=(\psi,u,h)^{\rm t}$ and compare the time-domain evolution with the waveform reconstructed from a superposition of QNMs in the frequency domain.
We first outline the computational methods employed in the time and frequency domains.

\subsection{Numerical methods}
\subsubsection{Time domain}
For the time-domain calculation, we solve the coupled wave equations directly using a method of lines scheme implemented in Julia.
In this approach, the spatial dependence is discretized first, while the resulting system remains continuous in time and can therefore be integrated as a system of ordinary differential equations.
We cover the numerical domain with a uniform grid, $r_{\ast,k}=-L+k\Delta r_\ast$ for $k=0,1,\cdots,N_x-1$, with $\Delta r_\ast=2L/(N_x-1)$, and approximate the second spatial derivative at each interior point with a second-order centered finite difference.
Denoting $\vec{\Psi}_k=\vec{\Psi}(t,r_{\ast,k})$ and $\vec{\Pi}_k=\partial_t\vec{\Psi}_k$, the discretized equations can be written directly as the first-order system
\begin{align}\label{eq:disc}
    \frac{d}{dt}\begin{pmatrix}\vec{\Psi}_k \\ \vec{\Pi}_k \end{pmatrix}
    =\begin{pmatrix} \vec{\Pi}_k \\ \frac{\vec{\Psi}_{k-1}-2\vec{\Psi}_k+\vec{\Psi}_{k+1}}{(\Delta r_{\ast})^2}-\bm{V}_k\vec{\Psi}_k \end{pmatrix}~,
\end{align}
where $\bm{V}_k=\bm{V}(r_k)$, with $k=1,2,\cdots,N_x-2$.

Equation~\eqref{eq:disc} is evaluated simultaneously at all interior grid points, thereby converting the original partial differential equations into one large system of ordinary
differential equations.  
We integrate this system with the adaptive fifth-order Runge--Kutta method~\texttt{Tsit5} from the Julia package~\texttt{OrdinaryDiffEq.jl}~\cite{DifferentialEquations.jl-2017}.  
Relative and absolute tolerances of $10^{-7}$ and $10^{-9}$ are imposed, respectively.  Although the internal time step is selected adaptively by the integrator, the numerical
solution is recorded at uniformly spaced output times.

As a simple choice of initial data at $t_0=0$, we set the perturbation variables to zero and prescribe their time derivatives (or equivalently, $\vec{\Pi}$) as normalized Gaussian profiles:
\begin{align}
    \vec{\Psi}(0,r_\ast)&=0~, \\
    \partial_t\vec{\Psi}(0,r_\ast)&=\vec{E}\frac{1}{\sqrt{2\pi}\sigma}\exp\left[-\frac{(r_\ast-r_\ast^{\rm src})^2}{2\sigma^2}\right]~.
\end{align}
Here, $r_\ast^{\rm src}$ and $\sigma$ denote the center and width of the initial pulse, while $\vec{E}$ specifies its amplitude in each component.
Homogeneous Dirichlet conditions are imposed at the two numerical boundaries. 
The boundaries are placed sufficiently far from the source and the observer so that reflected signals do not reach the observer during the time interval used for the waveform analysis.

\subsubsection{Frequency domain}

\begin{figure*}[t]
\centering
\includegraphics[width=0.9\linewidth]{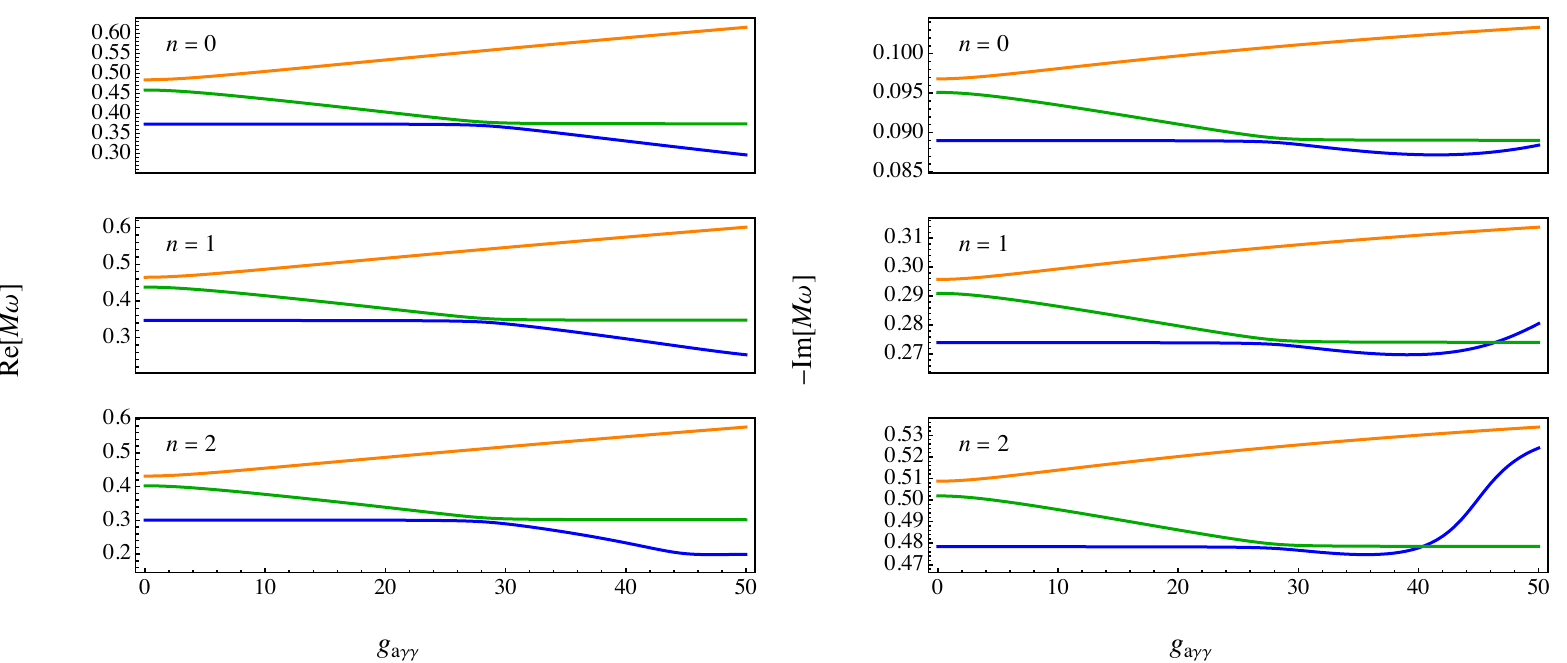}
\caption{QNM spectra for $n=0$, 1, and 2 as functions of the axion--photon coupling~$g_{a\gamma\gamma}$, for $Q/M=0.1$ and $l=2$. The blue, green, and orange curves represent the gravity-led, electromagnetic-led, and axion-led modes, respectively.}
\label{fig:QNM}
\end{figure*}

From the frequency-domain analysis in Sec.~\ref{sec:general}, the QNM contribution to the ringdown waveform is given by Eq.~\eqref{eq:Psi_t}.
To evaluate this expression, we need to obtain the QNM frequencies and the mode functions satisfying the boundary conditions in Eqs.~\eqref{eq:inmode} and \eqref{eq:upmode}.
We compute the QNM frequencies using the coupled-field implementation of Leaver's method~\cite{Leaver:1985ax,Leaver:1990zz} developed in our previous work~\cite{Takahashi:2025uwo}.\footnote{In Ref.~\cite{Takahashi:2025uwo}, we employed variables that diagonalize the gravitational and electromagnetic subsectors instead of the present variables~$(\psi,u,h)$. Note, however, that the QNM frequencies are independent of the choice of variables.}
Figure~\ref{fig:QNM} shows the QNM spectra for $n=0$, 1, and 2. We refer to the branches associated with the gravitational, electromagnetic, and axion degrees of freedom in the decoupling limit~$g_{a\gamma\gamma}\to0$ as the gravity-led, electromagnetic-led, and axion-led modes, respectively, and label them by $s=(g,e,a)$.

The mode functions are obtained by constructing asymptotic expansions near the horizon and at spatial infinity to impose the appropriate boundary conditions, and then numerically integrating the frequency-domain radial ordinary differential equations.
For the in-mode solutions, the near-horizon behavior is expanded as
\begin{align}
    \vec{\Psi}^{({\rm in},\alpha)}\sim(r-r_+)^{-\frac{i\omega r_+^2}{r_+-r_-}}\sum_{j=0}^{\infty}\vec{c}_j^{\ (\alpha)}(r-r_+)^j~,
\end{align}
where the leading coefficients are chosen as $\vec{c}_0^{\ (\alpha)}=\vec{\delta}_\alpha$.
For the up-mode solutions, the asymptotic behavior at spatial infinity is expanded as 
\begin{align}\label{eq:BC_up}
    \vec{\Psi}^{({\rm up},\alpha)}\sim r^{2iM\omega}e^{i\omega r}\sum_{j=0}^{\infty}\vec{d}_j^{\ (\alpha)}r^{-j}~,
\end{align}
where the leading coefficients are chosen as $\vec{d}_0^{\ (\alpha)}=\vec{\delta}_\alpha$.
Substituting these expansions into the radial equations, the coefficients~$\vec{c}_j^{\ (\alpha)}$ ($\vec{d}_j^{\ (\alpha)}$) for $j\geq1$ can be determined in terms of $\vec{c}_0^{\ (\alpha)}$ ($\vec{d}_0^{\ (\alpha)}$).
Using these asymptotic expansions as boundary conditions, we impose the boundary condition at $r=r_+(1+\epsilon)$ with $\epsilon\ll1$ for the in-mode solutions, and at a sufficiently large radius representing spatial infinity for the up-mode solutions.
The mode functions are then obtained by directly integrating the ordinary differential equations.
The frequency-domain calculations described above were performed using Mathematica.

Here, as the simplest setup, we consider the following delta-function initial data at $t_0=0$:
\begin{align}
    \vec{\Psi}(0,r_\ast)&=0~, \\
    \partial_t\vec{\Psi}(0,r_\ast)&=\vec{E}\ \delta(r_\ast-r_\ast^{\rm src})~.
\end{align}
This corresponds to the $\sigma\to0$ limit of the Gaussian-wave-packet initial data used in the time-domain calculation.
In this case, from Eq.~\eqref{eq:source}, the source term in the frequency domain is given by
\begin{align}
    \vec{S}=-\vec{E}\ \delta(r_\ast-r_\ast^{\rm src})~.
\end{align}
Substituting this into Eq.~\eqref{eq:Psi_t}, the QNM superposition describing the ringdown waveform is given by
\begin{align}\label{eq:Psi_f_delta}
    \vec{\Psi}(t,r)=\sum_{s,n}A_{s n}e^{-i\omega_{s n}t}\bm{\Psi}^{\rm (up)}\bm{T}(r_\ast^{\rm src})\vec{E}+{\rm (c.c.)}~.
\end{align}
The quantities~$A_{s n}$ and $\bm{T}$, which constitute the factor corresponding to the excitation coefficient in the single-field case, can be computed from the mode functions constructed above.

\subsection{Comparison of the results}

\begin{figure}[t]
\centering
\includegraphics[width=0.9\linewidth]{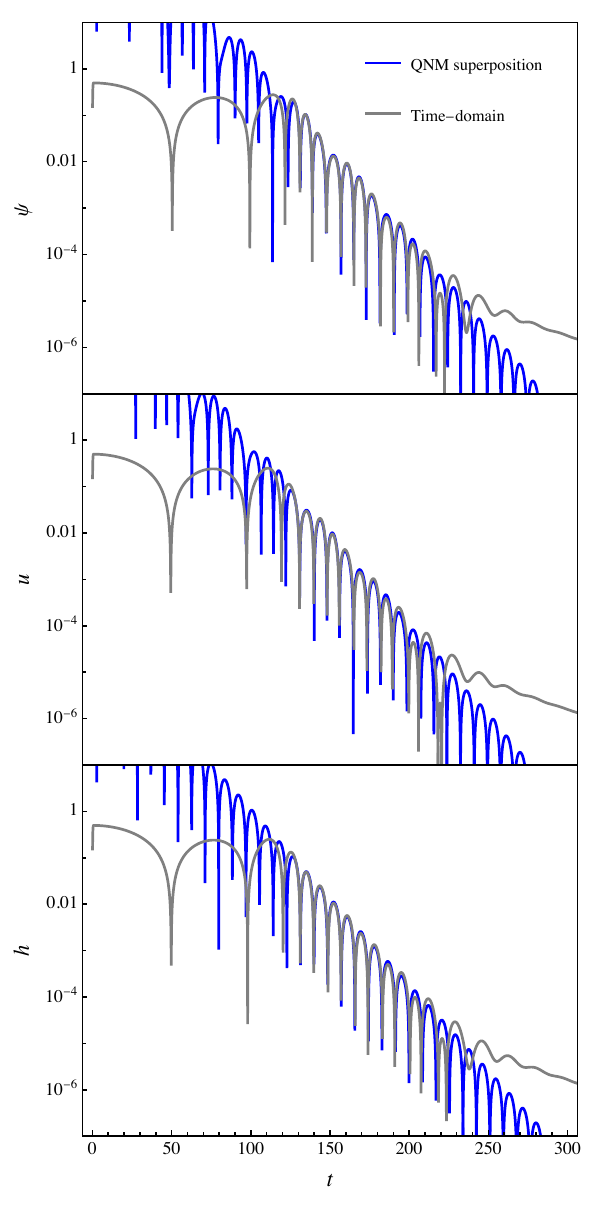}
\caption{Comparison of the ringdown waveforms obtained by direct time-domain integration (gray solid lines) and by the QNM superposition obtained in the frequency domain (blue solid lines). From top to bottom, the panels show the $\psi$, $u$, and $h$ components, respectively. The parameters are $Q=0.1$ and $g_{a\gamma\gamma}=28.2$, with the initial data amplitude~$\vec{E}=(1,1,1)^{\rm t}$. We set $M=1$.}
\label{fig:time_freq}
\end{figure}

We now compare the results obtained using the methods described above.
In our numerical implementation, we set $M=1$, and all quantities below are expressed in these units.
Figure~\ref{fig:time_freq} shows the waveform obtained from the time- and frequency-domain calculations for $l=2$, $Q=0.1$, and $g_{a\gamma\gamma}=28.2$.
The value~$g_{a\gamma\gamma}=28.2$ is chosen near the avoided crossing between the QNM branches associated with the gravitational and electromagnetic degrees of freedom, whose consequences will be discussed in the next section.
Here, we set the initial data amplitude to $\vec{E}=(1,1,1)^{\rm t}$.
The source is located at $r_{\ast}=r_\ast^{\rm src}=60$, and the waveforms are evaluated at $r_{\ast}=r_\ast^{\rm obs}=60$.
For the time-domain initial data, we set the width of the Gaussian profile to $\sigma=0.25$, which is chosen to be sufficiently small to approximate a delta function.
For the frequency-domain QNM superposition, we include only the fundamental mode associated with each degree of freedom, resulting in a superposition of three QNMs.

As shown in Fig.~\ref{fig:time_freq}, the direct time-domain evolution agrees well with the QNM superposition obtained in the frequency domain for $t\gtrsim r_\ast^{\rm src}+r_\ast^{\rm obs}=120$, until the late-time tail becomes significant.
This timescale corresponds to the time required for the initial Gaussian wave packet to propagate to the vicinity of the potential peak around $r_\ast=0$, and return to the observation point, in agreement with the intuitive expectation.
In particular, for each component, the two calculations agree not only in the oscillation frequencies but also in the amplitudes. This agreement validates the formalism developed in Sec.~\ref{sec:general} for systems with multiple coupled degrees of freedom.
For the QNM superposition obtained in the frequency domain, we do not use the approximation~$\bm{\Psi}^{\rm (up)}\sim e^{i\omega r_\ast}\bm{I}$ in Eq.~\eqref{eq:Psi_f_delta}, but instead use $\bm{\Psi}^{\rm (up)}$ constructed as in Eq.~\eqref{eq:Psi_matrix} from the mode functions obtained by imposing the boundary conditions as in Eq.~\eqref{eq:BC_up} and numerically integrating the perturbation equations. 
For comparison, for the present setup, the approximation~$\bm{\Psi}^{\rm (up)}\sim e^{i\omega r_\ast}\bm{I}$ leads only to a small discrepancy.

\section{Phenomenology}\label{sec:phenomenology}

\subsection{Ringdown waveforms and excitation coefficients}

\begin{figure*}[t]
\centering
\includegraphics[width=0.9\linewidth]{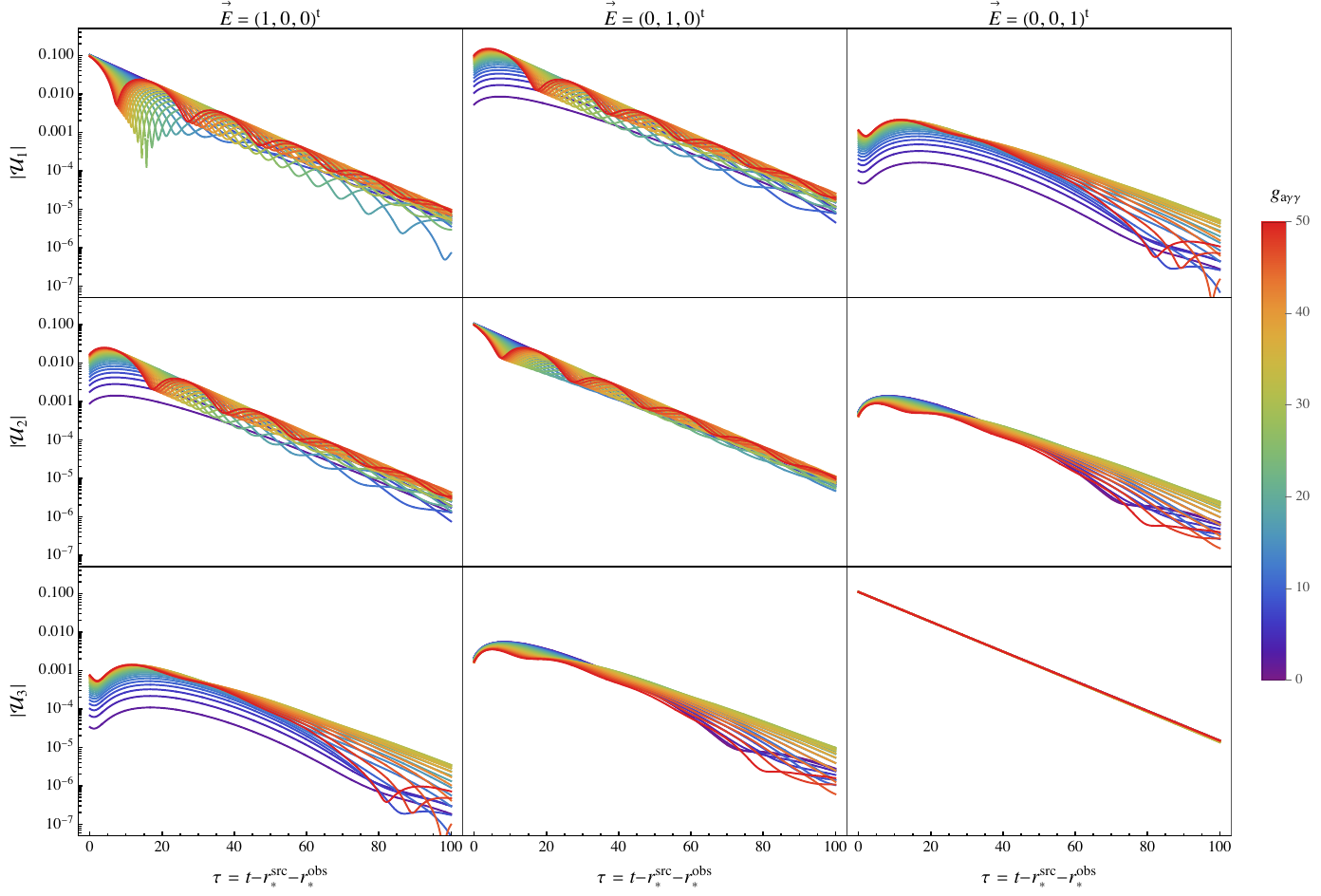}
\caption{Dependence on $g_{a\gamma\gamma}$ of the time evolution of the absolute values of the complex modes~$|{\cal U}_\alpha|$. From top to bottom, the panels show ${\cal U}_1$, ${\cal U}_2$, and ${\cal U}_3$, corresponding to the complex modes of $\psi$, $u$, and $h$, respectively. From left to right, the initial-data amplitudes are $\vec{E}=(1,0,0)^{\rm t}$, $(0,1,0)^{\rm t}$, and $(0,0,1)^{\rm t}$. The waveforms are constructed by superposing the three fundamental QNMs associated with the respective degrees of freedom, where the asymptotic form of $\bm{\Psi}^{\rm (up)}$ at spatial infinity is used. We set $M=1$ and $Q=0.1$.}
\label{fig:g_depndence}
\end{figure*}

Finally, we discuss the phenomenological consequences of avoided crossings in the QNM spectrum for the ringdown waveform of systems with multiple coupled degrees of freedom.
In particular, in the EMA system with $Q=0.1$, such an avoided crossing occurs around $g_{a\gamma\gamma}\simeq28.2$ between the QNM branches associated with the electromagnetic and gravitational degrees of freedom. Since this behavior can be interpreted as a resonant phenomenon, we examine how it affects the resulting ringdown waveform.
The parameter and source settings are the same as in the previous section, except for the axion--photon coupling~$g_{a\gamma\gamma}$ and the initial amplitudes.

As demonstrated in the previous section, the QNM superposition formulated in Sec.~\ref{sec:general} accurately captures the damped oscillatory part of the ringdown waveform. We therefore first use the QNM-reconstructed waveforms to investigate their dependence on the model parameters and initial amplitudes.
We write each component of the physical perturbation as
\begin{align}
    \Psi_\alpha={\cal U}_\alpha+{\rm (c.c.)}~,
\end{align}
where ${\cal U}_\alpha$ denotes the $\alpha$-th component of the first term on the right-hand side of Eq.~\eqref{eq:Psi_f_delta}, before adding its complex conjugate.
Here, assuming that the observer is located sufficiently far from the BH, we approximate $\bm{\Psi}^{\rm (up)}\sim e^{i\omega r_\ast}\bm{I}$, so that ${\cal U}_\alpha$ can be explicitly written as
\begin{align}\label{eq:complex_mode}
    {\cal U}_\alpha=\sum_{s}&A_{s 0}\left.\left(T_{\alpha 1}E_1+T_{\alpha 2}E_2+T_{\alpha 3}E_3\right)\right|_{\omega=\omega_{s 0},r_\ast=r_\ast^{\rm src}} \notag \\
    &\times e^{-i\omega_{s 0}(t-r_\ast)}~,
\end{align}
where only the $n=0$ contribution associated with each degree of freedom is included.
Figure~\ref{fig:g_depndence} shows the time evolution of the absolute value of the complex mode for each component for various values of the axion--photon coupling~$g_{a\gamma\gamma}$. 
These absolute values represent the envelopes of the corresponding physical waveforms up to an overall factor of two as $\Psi_\alpha=2\mathrm{Re}({\cal U}_\alpha)$.
We consider three choices of the initial amplitude, for which one component of $\vec{E}$ is set to unity while the others are set to zero.

In many cases shown in Fig.~\ref{fig:g_depndence}, the envelope itself exhibits an oscillatory behavior. 
This can be understood as a beating pattern arising from the superposition of damped oscillations with different frequencies, a common feature of systems with multiple degrees of freedom~\cite{Molina:2010fb,Melis:2024kfr}.
A notable feature, however, is that 
for certain combinations of waveform component and initial perturbation, the beating pattern disappears and the waveform decays more slowly in the vicinity of the avoided crossing.
This behavior is observed for ${\cal U}_1$ and ${\cal U}_2$ with $\vec{E}=(0,0,1)$ (top-right and middle-right panels, respectively), and for ${\cal U}_3$ with $\vec{E}=(1,0,0)$ and $\vec{E}=(0,1,0)$ (bottom-left and bottom-middle panels, respectively).
In particular, the dependence of the waveform behavior on $g_{a\gamma\gamma}$ is nonmonotonic, with the slowest decay occurring around $g_{a\gamma\gamma}=28.2$.
This behavior is expected from the single-field case.
As the two modes approach the exceptional point, their frequencies become nearly degenerate, resulting in an extremely long beating period.
Furthermore, near the exceptional point, suppose that the eigenfrequencies and coefficients can be expressed as $\omega_{j,k}=\omega_{\rm EP}(1\pm\delta)$ and $D_{j,k}=D_{\rm EP}(1\pm d/\delta)/2$, respectively.
For $|t\omega_{\rm EP}\delta|\ll1$, their superposition has approximately the same time dependence as a contribution from a double pole in the inverse Laplace transform~\cite{PanossoMacedo:2025xnf}:
\begin{align}
    D_j e^{-i\omega_j t}+D_k e^{-i\omega_k t}\simeq D_{\rm EP}(1-id\omega_{\rm EP}t)e^{-i\omega_{\rm EP}t}~.
\end{align}
The second term in parentheses is proportional to time, giving rise to an effectively slower decay.

\begin{figure}[t]
\centering
\includegraphics[width=1.0\linewidth]{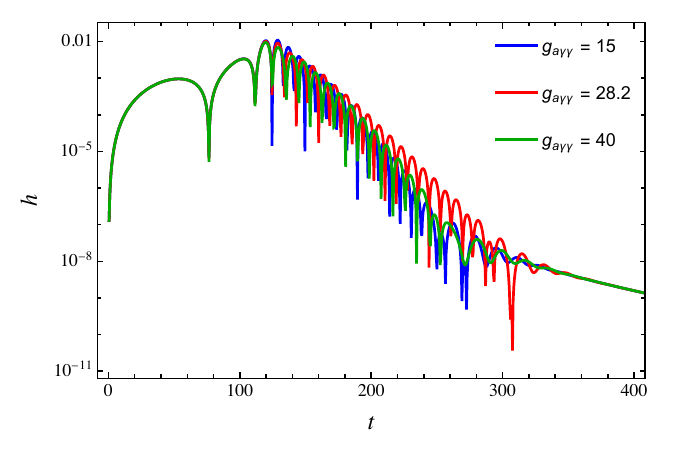}
\caption{Dependence of the gravitational perturbation waveform obtained by direct time-domain integration on $g_{a\gamma\gamma}$, with the initial amplitude~$\vec{E}=(0,1,0)$. The blue, red, and green curves correspond to $g_{a\gamma\gamma}=15$, $28.2$, and $40$, respectively. In particular, $g_{a\gamma\gamma}=28.2$ corresponds to the avoided crossing and exhibits a longer-lived ringdown than the other cases. We set $M=1$ and $Q=0.1$.}
\label{fig:h_comparison}
\end{figure}

We have also confirmed this behavior in the time-domain waveform.
Figure~\ref{fig:h_comparison} shows the gravitational waveforms for $g_{a\gamma\gamma}=15$, $28.2$, and $40$ with $\vec{E}=(0,1,0)$.
These correspond to the bottom-middle panel of Fig.~\ref{fig:g_depndence}.
Again, the waveform at $g_{a\gamma\gamma}=28.2$, where the avoided crossing occurs, exhibits the longest-lived ringdown.
This behavior cannot be inferred solely from the decay rates of the individual QNMs shown in Fig.~\ref{fig:QNM}.

\begin{figure*}[t]
\centering
\includegraphics[width=1.0\linewidth]{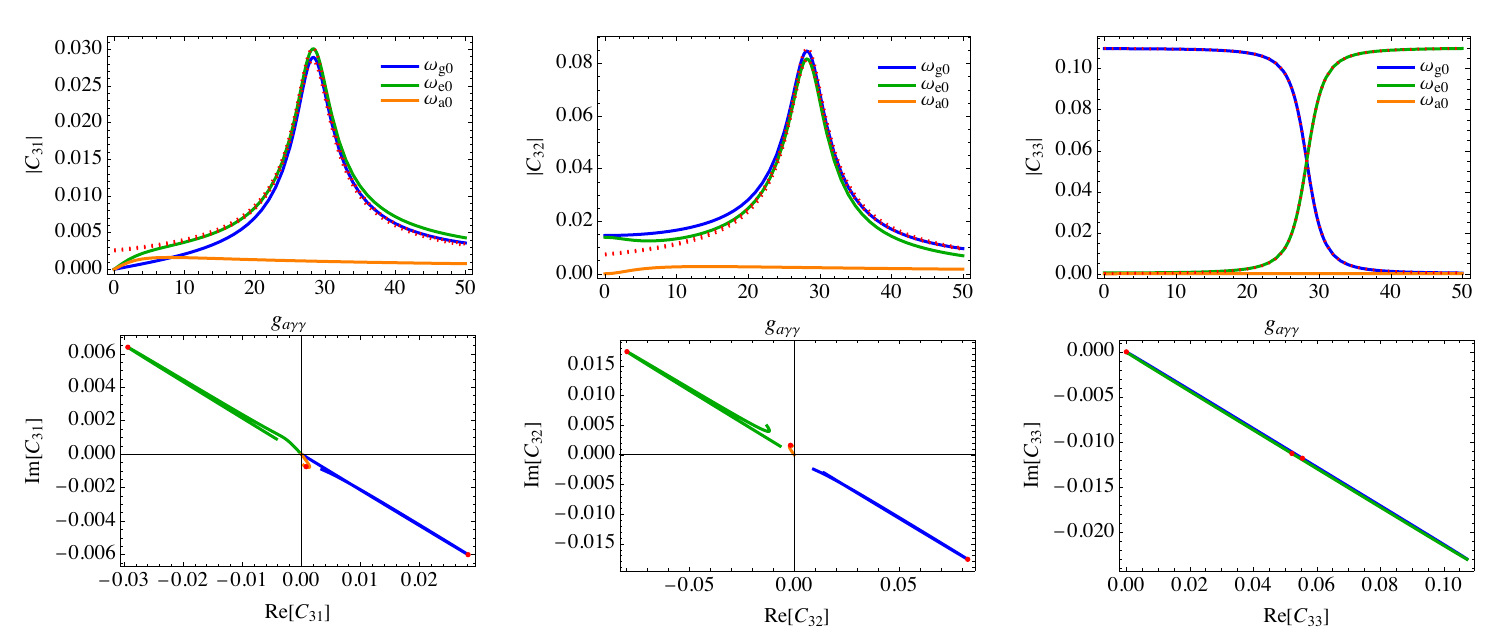}
\caption{Excitation coefficients of the gravitational component for different initial excitations as functions of $g_{a\gamma\gamma}$. The blue, green, and orange curves correspond to the gravity-led, electromagnetic-led, and axion-led QNMs, respectively. The upper panels show the absolute values of the excitation coefficients, while the lower panels show their trajectories in the complex plane. The red dotted curves in the upper panels represent fits based on the effective model given by Eqs.~\eqref{eq:Pdiag} and~\eqref{eq:Poffdiag}. The red dots in the lower panels indicate $g_{a\gamma\gamma}=28.2$. We set $M=1$ and $Q=0.1$.}
\label{fig:coefficients}
\end{figure*}

Again, as in the single-field case, despite the enhancement of the excitation factor~\cite{Takahashi:2025uwo}, no resonant enhancement is observed in the waveform itself.
In the multi-field case, however, the source integral involves mixing among the different field components through the factor~$\bm{T}$, and hence the cancellation responsible for the absence of resonant enhancement occurs in a nontrivial manner.
To elucidate this cancellation, we introduce the following quantity:
\begin{align}
    C_{\alpha\beta}^{(sn)}\coloneq A_{sn}\left.T_{\alpha\beta}\right|_{\omega=\omega_{sn},r_\ast=r_{\ast}^{\rm src}}e^{-i\omega_{sn}r_\ast^{\rm src}}~.
\end{align}
As can be seen from Eq.~\eqref{eq:complex_mode}, this quantity extracts, for each component~${\cal U}_\alpha$, the contribution proportional to the initial excitation~$E_\beta$, evaluated at $t=r_\ast^{\rm src}+r_\ast$.
It can thus be regarded as a generalized excitation coefficient that quantifies the excitation of each QNM by each component of the initial perturbation.
Figure~\ref{fig:coefficients} shows the excitation coefficients for the complex mode~${\cal U}_3$ of the gravitational perturbation~$h$.
As shown in the figure, the coefficients associated with $E_1$ and $E_2$ reach their maximum amplitudes around $g_{a\gamma\gamma}=28.2$, reflecting the resonant behavior at the avoided crossing. 
In the complex plane, however, the corresponding coefficients are enhanced in opposite directions, leading to a cancellation in the amplitude of the resulting waveform.

On the other hand, when only $E_3$ is initially excited, the waveform shows little dependence on the axion--photon coupling, as seen in the bottom-right panel of Fig.~\ref{fig:g_depndence}.
As can be seen from Eq.~\eqref{eq:h_pert}, the axion-photon coupling affects $h$ only indirectly through the electromagnetic sector, which provides a qualitative interpretation of the weak dependence observed in this setup.
However, an interesting feature emerges when the waveform is decomposed into individual QNM contributions.
As shown in Fig.~\ref{fig:QNM}, the gravity-led QNM eigenvalue remains nearly constant up to the vicinity of the avoided crossing at  $g_{a\gamma\gamma}\simeq28.2$, while after the avoided crossing, the electromagnetic-led mode becomes nearly constant.
The right panel of Fig.~\ref{fig:coefficients} shows that the dominant contributions of these two QNMs to the gravitational perturbation~$h$ interchange at the avoided crossing.
In other words, the physical characters of the two QNM branches are exchanged across the avoided crossing, so that no corresponding abrupt change appears when the system is viewed in terms of the gravitational perturbation itself.
This exchange of the physical characters of the QNMs can be interpreted as the mode-character exchange characteristic of a resonance phenomenon.

\subsection{Effective model}

The rapid variation of the excitation coefficients associated with the gravity- and electromagnetic-led modes near the resonance seen in Fig.~\ref{fig:coefficients} can be understood qualitatively from the same two-level model used to describe the avoided crossing~\cite{Motohashi:2024fwt,Takahashi:2025uwo}.
Following Ref.~\cite{Takahashi:2025uwo}, we write the two levels in the absence of their direct mixing as ${\cal E}_1=pe^{i\vartheta}$ and ${\cal E}_2=\omega_0^2e^{i\vartheta}$, corresponding to the electromagnetic- and gravity-led modes, respectively, and characterize their effective mixing by $qe^{i\vartheta}$.
Here, $p$, $q$, $\omega_0$, and $\vartheta$ are taken to be real, and $q$ is treated as approximately constant as $p$ is varied across the resonance.
In this model, the parameter~$p$ can be regarded as proportional to $g_{a\gamma\gamma}Q$~\cite{Takahashi:2025uwo}, so that increasing $g_{a\gamma\gamma}$ at fixed $Q/M$ corresponds to increasing $p$, with the resonance occurring around $p\simeq\omega_0^2$.

The two-level problem is thus described by the following effective Hamiltonian for the squared QNM frequencies:
\begin{align}
    \bm{H}_{\rm eff}
    \coloneq
    e^{i\vartheta}
    \begin{pmatrix}
        p & q \\
        q & \omega_0^2
    \end{pmatrix}~.
\end{align}
In terms of the resolvent~$(\omega^2\bm{I}-\bm{H}_{\rm eff})^{-1}$, with $\bm{I}$ being the $2\times 2$ identity matrix, the two QNM frequencies in this model correspond to poles, while the rapid parameter dependence of the excitation coefficients is encoded in the corresponding residues, up to factors associated with the initial data and normalization.
Denoting the eigenvalues of $\bm{H}_{\rm eff}$ by $\omega_\pm^2$, the residues are proportional to the matrices~$\bm{P}_\pm$, as can be read off from the identity
\begin{align}
    (\omega^2\bm{I}-\bm{H}_{\rm eff})^{-1}
    =\frac{\bm{P}_+}{\omega^2-\omega_+^2}+\frac{\bm{P}_-}{\omega^2-\omega_-^2}~,
\end{align}
with
\begin{align}
    \bm{P}_\pm\coloneq \frac{\bm{H}_{\rm eff}-\omega_\mp^2\bm{I}}{\omega_\pm^2-\omega_\mp^2}~.
\end{align}
More precisely, $\bm{P}_\pm$ are the residues with respect to $\omega^2$, whereas the residues with respect to $\omega$ at the QNM poles are
\begin{align}
    \underset{\omega=\omega_\pm}{\operatorname{Res}}\left(\omega^2\bm{I}-\bm{H}_{\rm eff}\right)^{-1}=\frac{\bm{P}_\pm}{2\omega_\pm}~.
\end{align}
The additional factor $1/(2\omega_\pm)$ varies slowly near the resonance and therefore does not affect the qualitative argument.
The diagonal components of $\bm{P}_\pm$ are given by
\begin{align}\label{eq:Pdiag}
    (\bm{P}_\pm)_{11}=1-(\bm{P}_\pm)_{22}
    =\frac{1}{2}\left[1\pm\frac{p-\omega_0^2}{\sqrt{(p-\omega_0^2)^2+4q^2}}\right]~,
\end{align}
while the off-diagonal components are given by
\begin{align}\label{eq:Poffdiag}
    (\bm{P}_\pm)_{12}=(\bm{P}_\pm)_{21}
    =\pm\frac{q}{\sqrt{(p-\omega_0^2)^2+4q^2}}~.
\end{align}

As $p$ increases across the resonance, one of the diagonal components~$(\bm{P}_\pm)_{22}$ decreases from a value close to $1$ to a value close to $0$, while the other increases from a value close to $0$ to a value close to $1$.
Meanwhile, the absolute values of the off-diagonal components exhibit a square-root Lorentzian profile peaked at $p=\omega_0^2$, as identified in Ref.~\cite{Takahashi:2025uwo}.
The rapid variation of $C_{33}^{(g0)}$ and $C_{33}^{(e0)}$ near the resonance is therefore qualitatively captured by the gravitational diagonal components~$(\bm{P}_\pm)_{22}$, while that of $C_{32}^{(g0)}$ and $C_{32}^{(e0)}$ is captured by the off-diagonal components~$(\bm{P}_\pm)_{21}=(\bm{P}_\pm)_{12}$.

These correspondences are explicitly shown in the upper panels of Fig.~\ref{fig:coefficients}.
In the upper-right panel, Eq.~\eqref{eq:Pdiag} is shown by the red dotted curve, while Eq.~\eqref{eq:Poffdiag} is shown by the red dotted curves in the upper-left and upper-middle panels.
Here, we identify $p$ with $g_{a\gamma\gamma}$ with a fixed $Q$, while $q$ and $\omega_0^2$ are determined by fitting $C_{33}^{(g0)}$.
The overall amplitudes are fitted separately for each case. 
The model shows good agreement with the numerical results for the excitation coefficients, particularly in the vicinity of the resonance.
Here, the correspondence is only qualitative, since $C_{\alpha\beta}^{(sn)}$ is defined in the original $(\psi,u,h)$ basis and also contains projection and normalization factors not included in the simple two-level model.
Therefore, the agreement for $C_{32}$ and $C_{33}$ is expected, whereas that for $C_{31}$ may be understood as an indirect manifestation of the resonance between the gravitational and electromagnetic fields through the coupling of the axion to the electromagnetic field.

\section{Conclusions}\label{sec:conc}
In this paper, we have investigated BH perturbation systems with multiple coupled degrees of freedom that exhibit avoided crossings in their QNM spectra and analyzed the resulting waveform behavior in both the time and frequency domains.
First, based on the formulation of Ref.~\cite{Takahashi:2025uwo}, we developed a general framework starting from the time-domain perturbation equations and derived, using the Green's matrix method, an expression for the waveform as a superposition of QNMs.
We then applied this framework to the EMA system as a concrete example that exhibits avoided crossings between QNMs associated with different degrees of freedom.
For this system, we computed the waveforms using Gaussian wave packet initial data in the time domain and the corresponding delta function initial data in the frequency domain.
In the regime where the waveform is expected to be well described by a superposition of QNMs, we found excellent agreement between the two calculations, including the amplitudes, thereby validating the formalism.

We also investigated how the avoided crossing in the QNM spectrum manifests itself in the physical waveform.
Examining the waveforms for various values of the axion--photon coupling in each of the gravitational, electromagnetic, and axion perturbation components, we found that in most cases they exhibit a beating pattern commonly seen in systems with multiple coupled degrees of freedom.
However, for certain components and choices of the initial amplitudes, the beating pattern disappears and the waveform exhibits a slower decay than those obtained for other parameter values.
This occurs at parameter values corresponding to the avoided crossing and can be understood, as in the single-field case, as arising from the superposition of modes with nearly degenerate frequencies, which gives rise to a longer-lived contribution characterized by a term proportional to time.
Even in the multi-field case, however, no resonant enhancement of the waveform amplitude was observed despite the enhancement of the excitation factors.
Compared with the single-field case, the cancellation of the resonant enhancement occurs in a nontrivial manner.
By examining the excitation coefficients introduced for each component of the initial perturbation, we confirmed that the relevant contributions are enhanced in opposite directions in the complex plane, resulting in their cancellation in the physical waveform.
Furthermore, even for initial excitations for which the longer-lived behavior does not appear, our analysis reveals an exchange of the physical characters of the QNM branches across the resonance.
This mode-character exchange may provide an interesting connection to phenomena commonly encountered in non-Hermitian physics.

Compared with the single-field case, an interesting feature of systems with multiple degrees of freedom is that resonances between QNMs associated with different degrees of freedom can occur for a range of overtone numbers.
With the accuracy of our present time-domain calculations, extracting individual overtone contributions was challenging.
Nevertheless, if similar signatures of slow decay associated with the avoided crossing can also be identified in the overtone modes, they may provide a characteristic signature of resonant mode mixing between different degrees of freedom.
We emphasize that the present analysis is a demonstration based on the EMA system, and it remains unclear to what extent the features found here are universal in generic systems with multiple coupled degrees of freedom.
In particular, it would be interesting to clarify which components exhibit the slower decay associated with an avoided crossing and how this behavior depends on the initial excitation.

The presence of an avoided crossing may also have important implications for ringdown analysis. As pointed out in Ref.~\cite{Imafuku:2026rpn}, particular care may be required when fitting such waveforms with a superposition of two QNMs.
At the same time, distinguishing the slow decay associated with the resonance from an ordinary superposition of two nearby QNMs directly from the waveform may itself be challenging.
Several approaches have been developed and investigated for extracting nearly degenerate QNMs from the ringdown signal~\cite{Morisaki:2025gyu,Iida:2026jox,Iida:2026ylf}.
A systematic investigation of these issues in more general multi-field systems, including the role of overtones and initial excitations, is left for future work.
Such studies will be important for identifying signatures of additional degrees of freedom in BH ringdown.


\acknowledgments{
This work is supported by JSPS (Japan Society for the Promotion of Science) KAKENHI Grant Nos.~JP25KJ0067 and JP25K17397 (T.T.), JP22K03639 and JP26K21813 (H.M.), and JP23K13101 (K.T.).
}


\bibliographystyle{mybibstyle}
\bibliography{bib}

\end{document}